\documentclass[prd,aastex,amsmath,amssymb,twocolumn,floatfix,aps,showpacs]{revtex4-1} 

\usepackage{graphicx}% Include figure files
\usepackage{dcolumn}% Align table columns on decimal point 
\usepackage{bm}% bold math

\usepackage{xcolor}
\definecolor{myblue}{rgb}{0.0, 0.0, 0.6}
\usepackage{hyperref}
\hypersetup {
  colorlinks = true,
  citecolor  = myblue,
  linkcolor  = myblue,
  urlcolor   = myblue
}

\begin{document}

\title{
  Corrections to the elastic proton-proton analyzing power parametrization at high energies
}%

\author{A.A. Poblaguev}\email{poblaguev@bnl.gov}
\affiliation{%
 Brookhaven National Laboratory, Upton, New York 11973, USA
}%

\date{December 16, 2019}% It is always \today, today,

\begin{abstract}
The HJET Polarized Atomic Hydrogen Gas Jet Target polarimeter (HJET) polarimeter was designed to measure the absolute polarization of the proton beams at the Relativistic Heavy Ion Collider. In these measurements, the small scattering angle elastic $pp$ single $A_N(t)$ and double $A_{NN}(t)$ spin analyzing powers can be precisely determined. The experimental accuracy achieved at HJET requires corrections to the $A_N(t)$ parametrization, conventionally used for such studies. In this paper we evaluate the corrections to the analyzing powers due to (i) the differences between the electromagnetic and hadronic form factors and (ii) the $m_p^2/s$ terms in the elastic spin-flip $pp$ electromagnetic amplitude. The corresponding alterations of the evaluated hadronic spin-flip amplitudes are about the same as the experimental uncertainties of the HJET measurements. The proposed corrections may have implications  for the elastic $pp$ forward real-to-imaginary amplitude ratios $\rho$ determined in unpolarized $pp$ experiments.

\end{abstract}

 
\maketitle

\section{Introduction}

\begin{figure}[t]
  \begin{center}
    \includegraphics[width=0.45\textwidth]{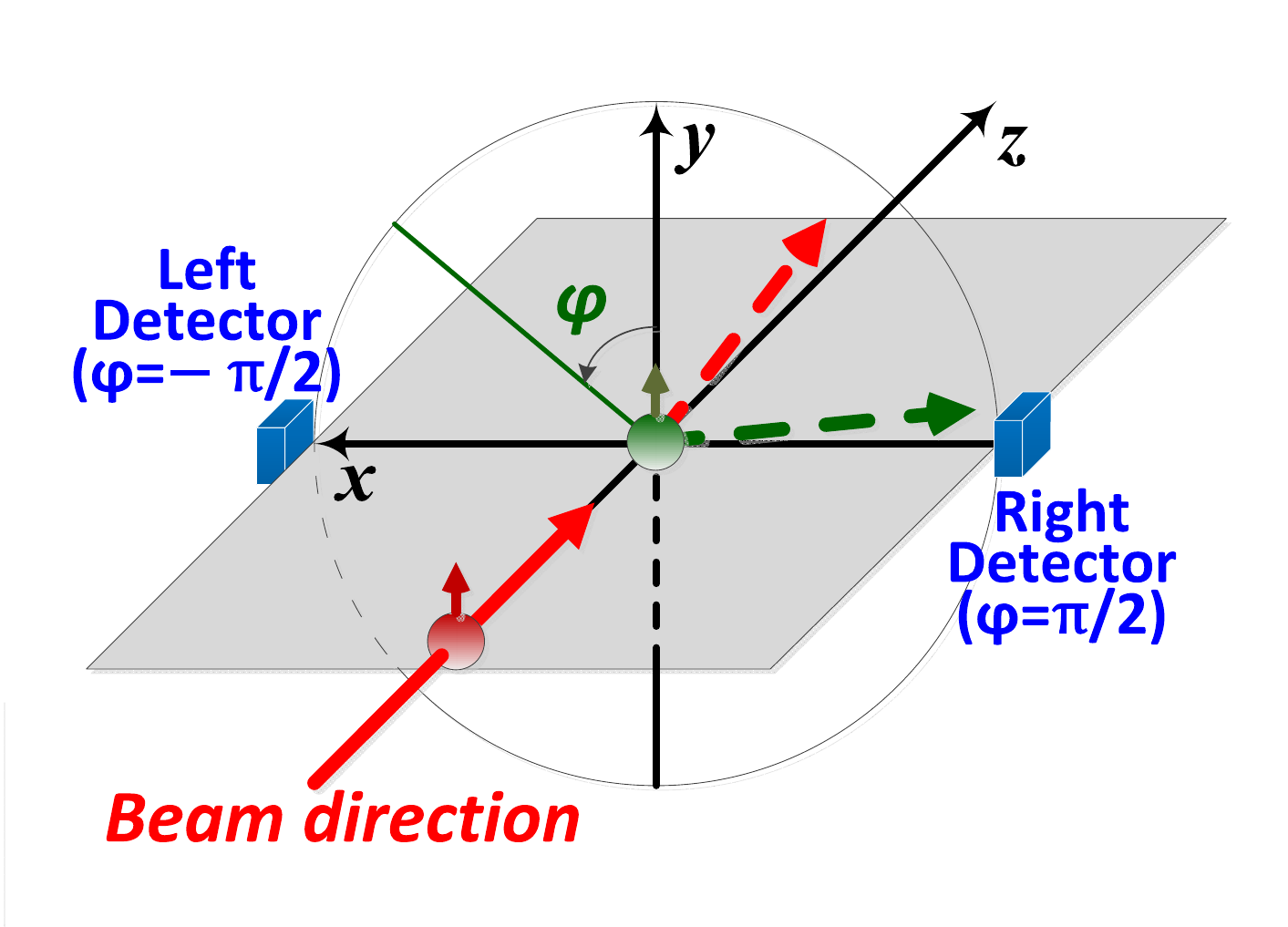}
    \caption{A schematic view of the $p^\uparrow p^\uparrow$ spin correlated asymmetry measurements at HJET. The recoil protons are counted in left/right symmetric detectors. The beam moves along the $z$ axis. The transverse polarization direction is along the $y$ axis.}
      \label{fig:Angles}
  \end{center}
\end{figure}

The Polarized Atomic Hydrogen Gas Jet Target\,\cite{bib:ABS} polarimeter (HJET) is employed to measure the absolute polarization of proton beams at the Relativistic Heavy Ion Collider (RHIC). For that, the vertically polarized proton beam is elastically scattered at small angles (Fig.\,\ref{fig:Angles}) on the vertically polarized target (the jet) with well-determined polarization $|P_j|=0.957\pm0.001$ and the beam and jet spin correlated asymmetries of the recoil protons [Eq.\,(\ref{eq:dsdt})] are studied.

The major upgrade of HJET in 2015, along with the development of new  methods in data analysis, allowed us to reduce the systematic uncertainties of the beam polarization measurements to a $\sigma^\text{syst}_P/P$\,$\lesssim$\,$0.5\%$\,\cite{bib:PSTP2017} level. Such a small systematic uncertainty of measurements, combined with large statistics of approximately $2\times10^9$ elastic $pp$ events per RHIC run accumulated in 2015 ($E_\text{lab}$\,=\,$100$\,GeV) and 2017 ($E_\text{lab}$\,=\,$255$\,GeV), allowed us to precisely measure the single $A_\text{N}$ and double $A_\text{NN}$ spin analyzing powers\,\cite{bib:Convention} in the Coulomb-nuclear interference (CNI) region.

Generally, $A_\text{N}(s,t)$ and $A_\text{NN}(s,t)$ are functions of the invariant variables $s$, center-of-mass energy squared, and $t$, 4-momentum transfer squared. An important part of the experimental study of the analyzing power is isolation of the hadronic spin-flip amplitudes. The theoretical basis for such studies was developed in Refs.\,\cite{bib:KL,bib:BGL}. An update\,\cite{bib:BKLST} for the RHIC spin program provided a parametrization of $A_\text{N}(s,t)$ which was used in all previous experimental evaluations\,\cite{bib:HJET06,bib:HJET09,bib:STAR13} of the hadronic spin-flip amplitudes in high energy near-forward elastic $pp$ scattering.

Recently it was pointed out\,\cite{bib:AbsorptiveCorr} that analyzing power $A_N(t)$ given in Ref.\,\cite{bib:BKLST} was derived with some simplifications, which might be essential for the experimental accuracy achieved at HJET:
{\em (i)} it was  implicitly assumed that the electromagnetic form factor is equal to the hadronic form factor $\exp(Bt/2)$, and
{\em (ii)} the elastic $pp$ electric form factor,  $G_E^{pp}$, was approximated, $G_E^{pp}$\!=\!$G_E^2(t)$, by an electric form factor $G_E(t)$ determined in electron-proton scattering experiments. The absorptive corrections, due to the initial and final state inelastic hadronic interactions between the colliding protons\,\cite{bib:AbsorptiveCorr}, were not considered in Ref.\,\cite{bib:BKLST}.

After this paper was accepted for publication, a theoretical evaluation of the absorptive corrections in elastic $pp$ scattering was given in Ref.\,\cite{bib:AbsorptiveCorr}. %However, these results are not discussed  in the paper.
These new results were not taken into account below.

Here, we analyze the effect of the possible corrections to the $A_\text{N}(t)$ parametrization on the results of the recent HJET measurements. The evaluated alteration of the measured hadronic spin-flip amplitudes suggests including the corrections in the data analysis. Also it was found that the discussed corrections may be important for experimental determination of the real-to-imaginary ratio $\rho$ of the $pp$ forward elastic scattering amplitude.

\section{Parametrization of the CNI analyzing powers at high energies}  

Elastic $p^\uparrow p^\uparrow$ scattering is described by five helicity amplitudes\,\cite{bib:BKLST}
\begin{eqnarray} 
\phi_1(s,t) &=& \langle ++ |M|++\rangle, \nonumber\\  
\phi_2(s,t) &=& \langle ++ |M|--\rangle, \nonumber\\  
\phi_3(s,t) &=& \langle +- |M|+-\rangle,          \\  
\phi_4(s,t) &=& \langle +- |M|-+\rangle, \nonumber\\  
\phi_5(s,t) &=& \langle ++ |M|+-\rangle. \nonumber
\label{eq:phi}   
\end{eqnarray}

For scattering in the CNI region,  the hadronic and electromagnetic components of the elastic $pp$ amplitude should be explicitly indicated,
\begin{equation}
  \phi_i = \phi_i^\text{h}+\phi_i^\text{em}\exp(i\delta_C).
\end{equation}
The Coulomb phase is approximately independent of helicity\,\cite{bib:BGL,bib:Cahn}
\begin{equation}
  \delta_C = \alpha\ln\frac{-2}{t\left(B+8/\Lambda^2\right)}-\alpha\gamma\:\:\sim\: 0.02,
  \label{eq:deltaC}
\end{equation}
where $\gamma$\,=\,$0.5772$ is Euler's constant and $\Lambda^2$\,=\,$0.71$\,GeV$^2$. For numerical estimates in Eq.\,(\ref{eq:deltaC}) and below, we assume the HJET measurement values of $s$ and $t$. The differential cross section slope  $B(s)$ depends on energy as $B_0+B_1\ln{s}$\,\cite{bib:Beznogikh} and is about $11.5$\,GeV$^{-2}$. To the lowest order in $\alpha$, the fine structure constant, the electromagnetic amplitudes were calculated in Ref.\,\cite{bib:BGL}.

For very low $t$, the hadronic amplitude is dominated by the
\begin{equation}
  \phi_+(s,t) = \left[\phi_1(s,t) + \phi_3(s,t)\right]/2
\end{equation}
term. According to the optical theorem,
\begin{equation}
  \text{Im\,}\phi_+^\text{h}(s,0) = \frac{\sigma_\text{tot}(s)\,s}{8\pi}\,\sqrt{1-4m_p^2/s},
\label{eq:phi+}
\end{equation}
where  $m_p$ is proton mass and $\sigma_\text{tot}(s)$ is the total $pp$ cross section.
Therefore, $\phi_+^\text{h}(s,t)$ can be presented as
\begin{equation}
  \phi_+^\text{h}(s,t) = (\rho+i)\,\frac{\alpha s}{-t_c}\left(1-4m_p^2/s\right)^{1/2}\displaystyle e^{Bt/2},
\end{equation}
where
\begin{eqnarray}
  \rho(s) &=& \text{Re\,}\phi_+^\text{h}(s,0)\,/\,\text{Im\,}\phi_+^\text{h}(s,0), \\
  t_c(s)  &=& -8\pi\alpha/\sigma_\text{tot}(s)\approx-1.84\times10^{-3}\,\text{GeV}^2, 
\end{eqnarray}
and $\exp{(Bt/2)}$ is the nuclear form factor. 

Similarly, hadronic single and double spin-flip amplitudes may be parametrized by the dimensionless factors
\begin{equation}
  r_5(s) = \frac{m_p\,\phi_5^\text{h}}{\sqrt{-t}\,\mbox{Im}\,\phi_+^\text{h}} = R_5+iI_5
\end{equation}
and
\begin{equation}
r_2(s) = \frac{\phi_2^\text{h}}{2\,\text{Im\,}\phi_+^\text{h}} = R_2+iI_2,
\end{equation}
respectively.

Using the expression for the elastic $pp$ cross section
\begin{equation}
\frac{d\sigma}{dt} = \frac{2\pi}{s(s\!-\!4m_p^2)}\,%
\left(  |\phi_1|^2\!+\!|\phi_2|^2\!+\!|\phi_3|^2\!+\!|\phi_4|^2\!+\!4|\phi_5|^2 \right),
\end{equation}
the single spin analyzing power can be presented\,\cite{bib:BKLST} as
\begin{eqnarray}
A_\text{N}(t) &=& \frac{-4\pi(d\sigma/dt)^{-1}}{s(s-4m_p^2)}\,%
\text{Im\,}\left[  \phi_5^*\left( \phi_1 + \phi_2 + \phi_3 - \phi_4 \right)\right]
\nonumber \\
   &=& \frac{\sqrt{-t}}{m_p}\,\, \frac{(t_c/t)\,f_\text{N}^0+f_\text{N}^1}{f_\text{cs}(t)},
  \label{eq:AN}
\end{eqnarray}
where
\begin{eqnarray}
f_\text{N}^0(r_5) & =& \varkappa\,(1-\rho\delta_C) -2(I_5-\delta_C R_5), \label{eq:f0_N}\\
f_\text{N}^1(r_5) & =& -2(R_5 -\rho I_5),  \label{eq:f1_N} \\
f_\text{cs}(t)~     & =& \left(\frac{t_c}{t}\right)^2-2(\rho+\delta_C)\frac{t_c}{t}+1+\rho^2. \label{eq:f_cs}
\end{eqnarray}
In Eq.\,(\ref{eq:f0_N}), $\varkappa$\,=\,$\mu_p-1$\,=\,$1.793$ is the proton's anomalous magnetic moment. 
The $A_N(t)$ dependence on $r_5$ appears in a linear function of $t$
\begin{equation}
  f_\text{N}(t,r_5) = f^0_\text{N}+f^1_\text{N}t/t_c \approx \varkappa - 2I_5 - 2R_5\,t/t_c
  \label{eq:lin}
\end{equation}
while the dependence on $r_2$ is negligible.

Similarly\,\cite{bib:BKLST},
\begin{eqnarray}
A_\text{NN}(t) &=& \frac{4\pi(d\sigma/dt)^{-1}}{s(s-4m^2)}\,%
\left[  2|\phi_5|^2 + \text{Re\,}\left(\phi_1\phi_2^*-\phi_3\phi_4^* \right)\right] 
\nonumber \\
 & = & \frac{(t_c/t)\,f_\text{NN}^0+f_\text{NN}^1}{f_\text{cs}(t)},  \label{eq:ANN} \\
f_\text{NN}^0(r_2)   & =&  -2(R_2 + \delta_CI_2),   \\
 f_\text{NN}^1(r_2) & =&  2I_2 + 2\rho R_2 - 
 ( \rho\varkappa-4R_5)\frac{\varkappa t_c}{2m_p^2}. \label{eq:f1_NN}
\end{eqnarray}

\section{Analyzing Power measurements at HJET}

For elastic scattering of vertically polarized beam and target protons, the recoil proton azimuthal angle $\varphi$ distribution is given\,\cite{bib:Convention} by
\begin{eqnarray}
\frac{d^2\sigma}{dtd\varphi} &=& \frac{1}{2\pi}\frac{d\sigma}{dt}\times%
\big[1+A_\text{N}\sin{\varphi}\left(P_j+P_b\right) +   \nonumber  \\ 
& &~~~\left(A_\text{NN}\sin^2{\varphi}+A_\text{SS}\cos^2{\varphi}\right)P_bP_j  \big].
\label{eq:dsdt}
\end{eqnarray}
Here, $\varphi$ used  is defined in accordance with Fig.\,\ref{fig:Angles}, and $P_j$ and $P_b$ are jet and beam polarizations, respectively.

 For HJET detectors, $\sin{\varphi}$\,=\,$\pm1$ and, thus, three spin correlated asymmetries $A_\text{N}P_{j}$, $A_\text{N}P_{b}$, and $A_\text{NN}P_jP_b$ can be experimentally determined in the momentum transfer range $0.001\lesssim -t \lesssim 0.020\,\text{GeV}^2$. Consequently, one can derive the beam polarization $P_b$ (the main purpose of HJET) as well as analyzing powers $A_\text{N}(t)$ and $A_\text{NN}(t)$.

The preliminary analysis of the HJET data acquired in RHIC runs 2015 and 2017 has been done using the analyzing power formulas of Ref.\,\cite{bib:BKLST}. The values of $\sigma_\text{tot}(s)$ and $\rho(s)$ were taken from Ref.\,\cite{bib:Menon} fit. The slope $B(s)$ was derived from Ref.\,\cite{bib:Bartenev}. Only for numerical estimates below, these preliminary results could be summarized as

\noindent{Run\,15\;(100\,GeV):}\quad$\sqrt{s}$\,=\,$13.76$\,GeV,\\
\hspace*{1em}$\rho$\,=\,$-0.079$, $\sigma_\text{tot}$\,=\,$38.39$\,mb, $B$\,=\,$11.2\pm0.2$\,GeV$^{-2}$,\\
 \hspace*{1em}$R_5  =   \left(  -15.5 \pm 0.9_\text{stat}\pm 1.0_\text{syst}  \right)\times10^{-3}$, \\
 \hspace*{1em}$I_5\;=   \left(    -0.7 \pm 2.9_\text{stat}\pm 3.5_\text{syst} \right)\times10^{-3}$, \\
 \hspace*{1em}$R_2  =  \left(  -3.65 \pm 0.28_\text{stat}                    \right)\times10^{-3}$, \\
 \hspace*{1em}$I_2\;=   \left( -0.10 \pm 0.12_\text{stat}                    \right)\times10^{-3}$. \\
\noindent{Run\,17\;(255\,GeV):}\quad$\sqrt{s}$\,=\,$21.92$\,GeV,\\ 
  \hspace*{1em}$\rho$\,=\,$-0.009$, $\sigma_\text{tot}$\,=\,$39.19$\,mb, $B$\,=\,$11.6\pm0.2$\,GeV$^{-2}$, \\
  \hspace*{1em}$R_5  =   \left(  -7.3 \pm 0.5_\text{stat}\pm 0.8_\text{syst} \right)\times10^{-3}$, \\
  \hspace*{1em}$I_5\;=   \left( \;21.5 \pm 2.5_\text{stat}\pm 2.5_\text{syst} \right)\times10^{-3}$, \\
  \hspace*{1em}$R_2  =  \left(  -2.15 \pm 0.20_\text{stat}                  \right)\times10^{-3}$, \\
  \hspace*{1em}$I_2\;=   \left( -0.35 \pm 0.07_\text{stat}                  \right)\times10^{-3}$. \\
For $r_2$, systematic errors are small in these measurements.

\section{Corrections to the analyzing powers}

To calculate corrections to Eq.\,(\ref{eq:AN}), it is convenient to use the scaled amplitudes
\begin{equation}
  \varphi_i(s,t) =\phi_i(s,t) / \text{Im\,}\phi_+^\text{h}(s,t).
\end{equation}
Since a possible dependence of $\rho$, $r_2$, and $r_5$ on $t$ may be neglected in the CNI region, the scaled hadronic amplitudes can be approximated by
\begin{equation}
  \begin{aligned}
    \varphi_1^\text{h} =  \varphi_3^\text{h} &=  \rho(s)+i, \\  
    \varphi_2^\text{h} &=  2r_2(s),  \\
    \varphi_4^\text{h} &=  r_4(s)\times(-t/m_p^2)\approx0, \\
    \varphi_5^\text{h} &= r_5(s)\times\sqrt{-t}/m_p.
  \end{aligned}
\end{equation}

For the electromagnetic amplitudes, we should include the corrections of order of $m_p^2/s$ which can be significant for $E_\text{Lab}$\,=\,$100$\,GeV. Using the following expressions for the proton's electromagnetic form factors\,\cite{bib:Sachs,bib:BKLST}, 
\begin{equation}
F_1=\frac{G_E-G_Mt/4m_p^2}{1-t/4m_p^2},\quad\varkappa F_2=\frac{G_M-G_E}{1-t/4m_p^2}
\end{equation}
and neglecting the $t/s$ terms, one can derive from Ref.\,\cite{bib:BGL}
\begin{eqnarray} 
   \varphi_1^\text{em} = \varphi_3^\text{em} &=& \varphi_0^\text{em}%
   \times(1\!-\!2m_p^2/s)/\sqrt{1\!-\!4m_p^2/s},\hspace*{2.5em} \nonumber \\  
  -\varphi_2^\text{em} = \varphi_4^\text{em} &=& \varphi_0^\text{em} F_\varkappa^2\!\times\!%
  (1\!-\!2m_p^2/s)/\sqrt{1\!-\!4m_p^2/s}, \label{eq:em} \\
   \varphi_5^\text{em} &=& \varphi_0^\text{em}\,\left(F_\varkappa-\frac{\sqrt{-t}}{2m_p}\,\frac{2m_p^2}{s-4m_p^2}\right). \nonumber 
\end{eqnarray} 
The following shorthand was used
\begin{eqnarray}
  \varphi_0^\text{em}    &=& \frac{t_c(s)}{t}\times F_1^2(t)\exp(-Bt/2) \\
  F_1~    &=&\frac{1-\mu_pt/4m_p^2}{1-t/4m_p^2}\times\left( 1+r_E^2t/6\right), \label{eq:F_1} \\
  F_\varkappa&=&\frac{\sqrt{-t}}{2m_p}\,\frac{\varkappa F_2\,}{F_1}%
  =\frac{\sqrt{-t}}{2m_p}\,\frac{\varkappa}{1\!-\!\mu_pt/4m_p^2}.  \label{eq:Fmu} 
\end{eqnarray}

The proton's  electric form factor $G_E(t)$ was approximated in (\ref{eq:F_1}) by the proton charge radius $r_E$\,=\,$\langle r_E^2\rangle^{1/2}$. In Eq.\,(\ref{eq:Fmu}), we did not distinguish between proton electric $r_E$ and magnetic $r_M$ radii.

The electromagnetic and hadronic form factors difference can be realized by the  substitution $t_c$\,$\rightarrow$\,$t_c+bt$ where
\begin{equation} 
  b/t_c = \frac{d}{dt}\left[F_1^2(t)e^{-Bt/2}\right]_{t=0}.
 \label{eq:b}
\end{equation}
Since the electric form factor $G_E(t)$ in the dipole form\,\cite{bib:GD,bib:Cahn} 
\begin{equation}
G_D(t)~ = \left(1-t/\Lambda^2\right)^{-2}, ~~~\Lambda^2=0.71\,\text{GeV}^2  \label{eq:GD} 
\end{equation}
was commonly used in the elastic $pp$ data analysis, it is convenient to explicitly isolate the corresponding term $b_D$ in (\ref{eq:b})
\begin{equation}
  b = b_D+b_\text{nf},
\end{equation}
where
\begin{equation}
  b_D/t_c = \frac{d}{dt}\left(G^2_D(t)e^{-Bt/2}\right)\big|_{t=0} = \left(\frac{4}{\Lambda^2}-\frac{B}{2}\right).  \label{eq:b_D}
\end{equation}
For the RHIC beam energies, 
\begin{eqnarray}
\text{100\,GeV:}~~~b_D &=&  \left(-0.06\pm0.19\right)\times10^{-3}, \label{eq:bD100} \\
\text{255\,GeV:}~~~b_D &=&  \left(+0.31\pm0.19\right)\times10^{-3}. \label{eq:bD255}
\end{eqnarray} 
The errors here correspond to the  systematic uncertainties in the values of $B(s)$\,\cite{bib:Bartenev}.

For $b_\text{nf}$ one finds
\begin{equation}
  b_\text{nf}/t_c = r_E^2/3-4/\Lambda^2-\varkappa/2m_p^2.
\end{equation}

Currently, PDG\,\cite{bib:PDG} gives two values of proton charge radius
\begin{eqnarray}
  r_{ep}     &=& 0.8751\pm0.0061\,\text{fm},\\
  r_{\mu p} &=& 0.84086\pm0.00026\pm0.00029\,\text{fm}, 
\end{eqnarray}
obtained in three kinds of measurements: with atomic hydrogen, with electron scattering off hydrogen, and with muonic hydrogen. 
The discrepancy between the methods is not resolved yet. Assuming $r_E$\,=\,$0.858\pm0.017$\,fm,  one obtains 
\begin{equation}
  b_\text{nf} = \left(0.64\pm0.46\right)\times10^{-3}. 
\end{equation}

Approximating $\varphi_0^\text{em}$\,=\,$(t_c/t)e^{bt/t_c}$, one finds a correction to the denominator (\ref{eq:f_cs}) of the analyzing power expressions
\begin{eqnarray}
    f_\text{cs}(t,\rho) 
    & \rightarrow & f_\text{cs}(t,\rho-b+\varkappa^2t_c/4m_p^2) \nonumber \\
    & + & b^2-2b\delta_C -2\rho\varkappa^2t_c/4m_p^2 +\dots \nonumber\\
    &\approx&  f_\text{cs}(t,\rho-b_D-b_\text{cs}). \label{eq:f_cs_eff}
\end{eqnarray}
where 
\begin{equation}
  b_\text{cs}=b_\text{nf}-\varkappa^2t_c/4m_p^2 
  = (2.3\pm0.5)\times10^{-3}.   \label{eq:b_cs}
\end{equation}
The term $\varkappa^2t_c/4m_p^2$ here is due to the spin-flip amplitude $\varphi_5^\text{em}$ contribution to $d\sigma/dt$.

The experimental determination of the real to imaginary ratio $\rho$ at high energies is based on an analysis of the $d\sigma/dt\,(t) \propto f_\text{cs}(t,\rho)$.  The proton-proton electromagnetic form factor was approximated by $G_D^2(t)$ in almost all experimental studies of $\rho$. Therefore, a biased value of $\rho$ was measured in these experiments,  $\rho^\text{exp}$\,=\,$\rho -b_\text{cs}$. The bias is small compared to the uncertainty of measurements in any of the experiments listed in PDG, but it may be substantial for the global fit\,\cite{bib:COMPETE}.
Since the  values of $\rho$  from the global fit are used in the analyzing power measurements,  we should replace
\begin{equation}
  \rho \rightarrow \rho + b_\text{cs}  \label{eq:rho_corr}
\end{equation}
in (\ref{eq:f_cs_eff}) as well as in the expressions for $f_\text{N}^0$, $f_\text{N}^1$, and $f_\text{NN}^1$ above.   Thus, the  leading order corrections to the analyzing power $A_N(t)$ from  Ref.\,\cite{bib:BKLST}   can be approximated in (\ref{eq:AN}) as 
\begin{eqnarray}
 f_\text{cs}(t,\rho)            &\to& f_\text{cs}(t,\rho-b_D) \label{eq:f}, \\
  f^0_\text{N} &\to& f^0_\text{N} - 2m_p^2/s   \label{eq:f0},\\
  f^1_\text{N} &\to& f^1_\text{N} + \varkappa\,\left(b_D+b_\text{nf}+b_\varkappa\right) \label{eq:f1},
\end{eqnarray}
where
\begin{equation}
  b_\varkappa = \mu_p t_c%
  \left[\frac{1}{4m_p^2}\,+\!\frac{r_M^2\!-\!r_E^2}{6\varkappa}\right]\approx\left(-1.4\!\pm\!0.7\right)\!\times\!10^{-3}
\end{equation}
reflects the spin-flip contribution [see Eq.\,(\ref{eq:Fmu})] to the electromagnetic form factor. The specified error is dominated by the  
experimental uncertainties in the value of proton magnetic radius $r_M=0.851\!\pm\!0.026\,\text{fm}$\,\cite{bib:rM}.

For $A_\text{NN}(t)$, the corrections are small compared to uncertainties of the measurement at HJET. Also, we can neglect the correction to the Coulomb phase $\delta_C(r_E,B)$.

\section{Numerical estimates of the corrections}

\begin{figure}[t]
  \begin{center}
    \includegraphics[width=0.47\textwidth]{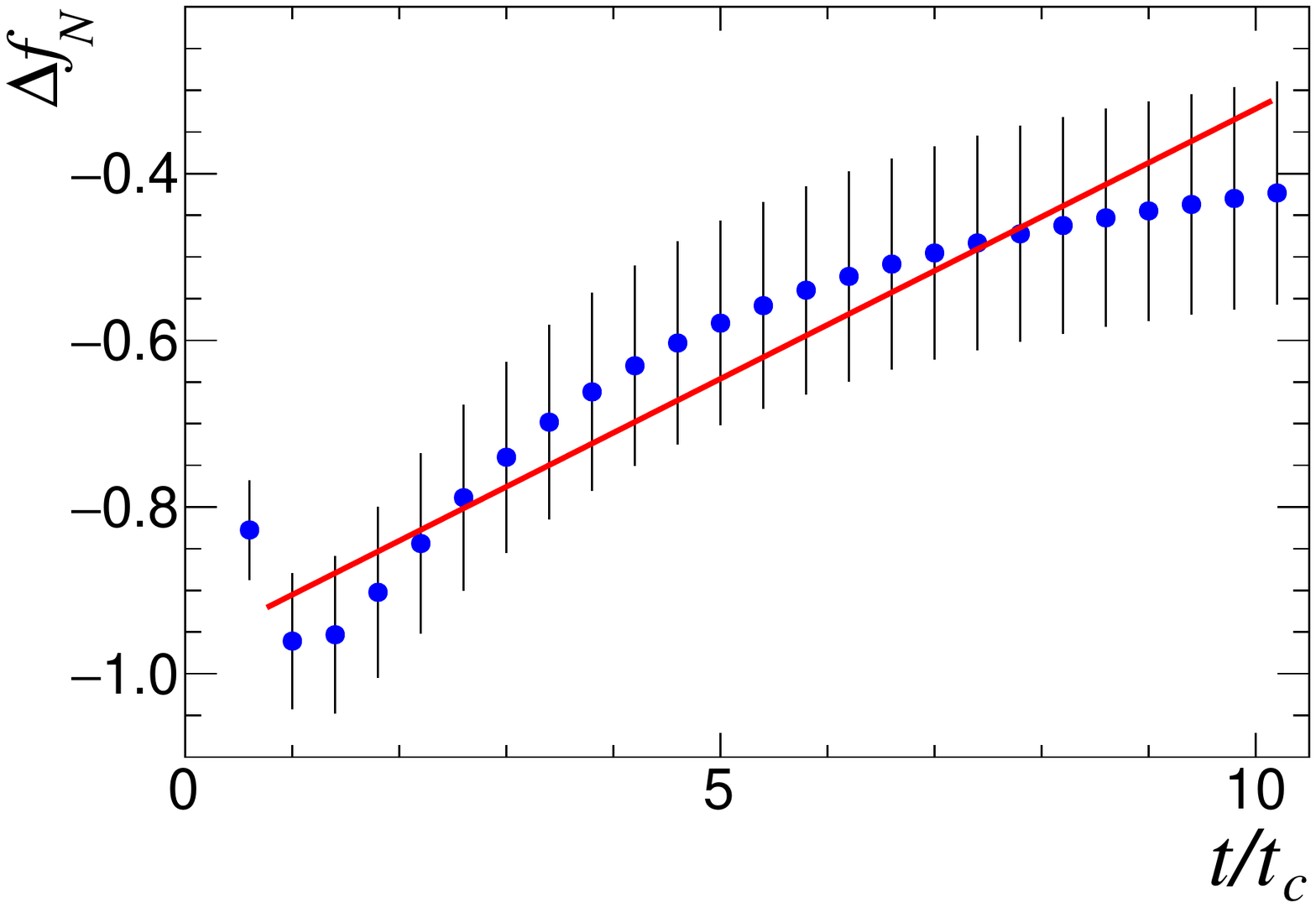}
    \caption{Calculation of the correction function $\Delta f_\text{N}(t)$  (blue points). The displayed error bars $\sigma_{0.1}(t)$ correspond to the HJET measurement statistical uncertainties if $b_D$\,=\,$0.1$. The error dependence on $b_D$ can be approximated by $\sigma(t,b_D)$\,=\,$\sigma_{0.1}(t)\times 0.1/b_D$. The red line is a linear fit. }
    \label{fig:bE}
  \end{center}
\end{figure}

The effect of the substitution (\ref{eq:f}) can be parametrized by the effective correction $b_D\Delta f_\text{N}(t)$ to the linear function $f_\text{N}(t)$
\begin{equation}
 \frac{1}{ f_\text{cs}(t,\rho-b_D)}=\frac{1+b_D\Delta f_\text{N}(t)}{f_\text{cs}(t,\rho)}. 
\end{equation}
The dependence of $\Delta f_\text{N}(t)$ on $\rho$ and $b_D$ can be neglected.  The calculated value of this correction is shown in Fig.\,\ref{fig:bE}. For the HJET values of $b_D$ given in Eqs.\,(\ref{eq:bD100}) and (\ref{eq:bD255}), the nonlinearity is not experimentally observable, and, thus,  we can approximate
\begin{equation}
\Delta f_\text{N}(t) = c_0 + c_1\,t/t_c
\end{equation}
or, equivalently,
\begin{equation}
  f^i_\text{N} \to f^i_\text{N} + c_i\varkappa b_D,~~i=0,1. \label{eq:fc}
\end{equation}
Obviously, the values of $c_0$ and $c_1$ depend on the $t$-range and experimental uncertainties. The HJET data analysis leads to $c_0$\,$\sim$\,$-1.0$ and $c_1$\,$\sim$\,$0.1$.

\begin{figure}[t]
  \begin{center}
    \includegraphics[width=0.47\textwidth]{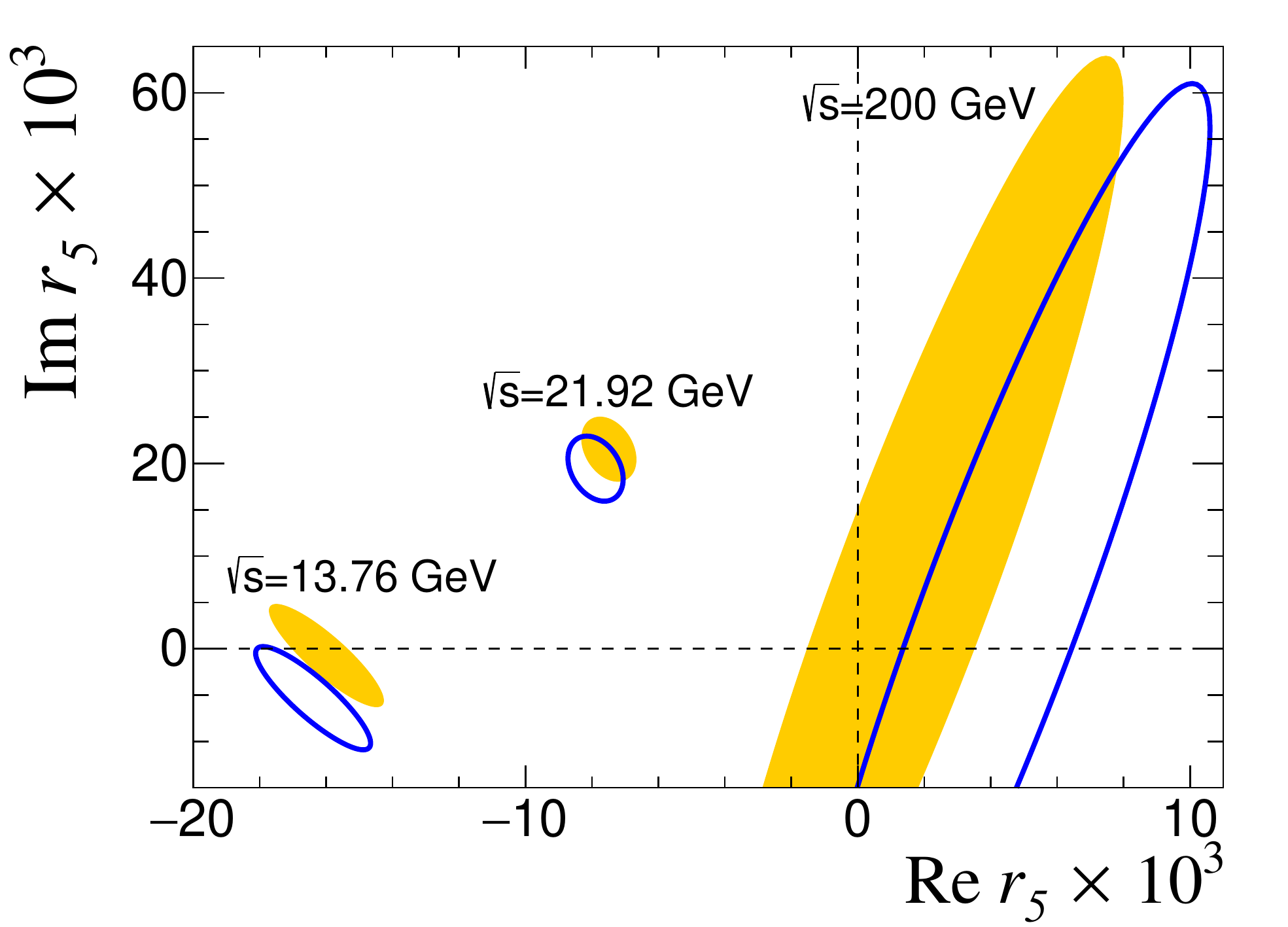}
    \caption{$\Delta\chi^2$\,=\,$1$ correlation (stat+syst) contours for $r_5$ with (solid lines) and without (filled areas) corrections to $A_\text{N}$. The absorptive corrections are not included. To display the $\sqrt{s}$\,=\,$200\,\text{GeV}$ contours, we used the data from Ref.\,\cite{bib:STAR13}.}
    \label{fig:r5}
  \end{center}
\end{figure}

Combining (\ref{eq:f0}), (\ref{eq:f1}), and (\ref{eq:fc}) we find the corrections to the measured hadronic form factors as follows:
\begin{eqnarray}
  \Delta I_5 &=& (\varkappa/2)\times c_0b_D   -   m_p^2/s, \label{eq:dI5} \\
  \Delta R_5 &=& (\varkappa/2)\times\left[(1+c_1)b_D+b_\text{nf}+b_\varkappa\right] + \rho\Delta I_5. \label{eq:dR5}
\end{eqnarray}
For HJET measurements, the calculation gives
\begin{equation}
\begin{aligned}[t]
  \text{100\,GeV:}~&\Delta R_5\!=\! \left( -0.4\!\pm\!0.2_B\!\pm\!0.4_{r_E}\!\pm\!0.6_{r_M} \right)\!\times\!10^{-3}, \\
                                         &\Delta I_5 = \left( -4.6\!\pm\!0.2_B \right)\!\times\!10^{-3}, \\
  \text{255\,GeV:}~&\Delta R_5\!=\!\left( -0.4\!\pm\!0.2_B\!\pm\!0.4_{r_E}\!\pm\!0.6_{r_M} \right)\!\times\!10^{-3}, \\
                                         &\Delta I_5 = \left( -2.1\!\pm\!0.2_B \right)\!\times\!10^{-3}.  \label{eq:dI5R5}
\end{aligned}
\end{equation}
The errors here are due to uncertainties in values of $B$, $r_E$, and $r_M$. Each error is strongly correlated through Eqs.\,(\ref{eq:dI5R5}). The large corrections to $I_5$ are due to the term $m_p^2/s$ in (\ref{eq:dI5}), which is 0.0047 for 100 GeV and 0.0018 for 255 GeV. Alterations of the measured $r_5$ are comparable with the experimental uncertainties (see Fig.\,\ref{fig:r5}) and, thus, should not be neglected.

\section{Possible effect of the absorptive correction}

A dependence of the measured $r_5$ on the absorptive corrections could be readily estimated if the corresponding modification of the  electromagnetic form factor $\mathcal{F}^\text{em}(t)$
of an elastic $pp$ amplitude can be approximated in the CNI region by a linear function of $t$,
\begin{equation}
  \mathcal{F}^\text{em}(t) \rightarrow   \mathcal{F}^\text{em}(t)\times\left[1+a(s)t/t_c\right]. \label{eq:abs}
\end{equation}

Generally, $a(s)$ is spin dependent. It can be effected by the substitutions
$b_\text{nf} \to b_\text{nf}$+$a_\text{nf}$,
$b_\varkappa \to b_\varkappa$+$a_\text{sf}$$-$$a_\text{nf}$,
where $a_\text{nf}(s)$ and $a_\text{sf}(s)$ are absorptive corrections to nonflip and spin-flip amplitudes, respectively.
The dominant absorptive corrections to $r_5$ and $r_2$ can be written as  
\begin{eqnarray}
\Delta_\text{a}R_5 = a_\text{sf}\varkappa/2,\phantom{0}
&~~~~&
\Delta_\text{a}I_5 = -a_\text{nf}\delta_C\varkappa/2\approx0,
\label{eq:dR5a} \\
\Delta_\text{a}R_2 = 0,\phantom{a_\text{sf}\varkappa/2}
&~~~~&
\Delta_\text{a}I_2  =  a_\text{nf}\frac{\varkappa^2t_c}{4m_p^2}\approx0. 
\label{eq:dI2a} 
\end{eqnarray}
As it was underlined above, the correction, such as given in Eq.\,(\ref{eq:abs}), does not modify $f_\text{cs}(t,\rho)$ but specifies the systematic errors in the experimental determinations of $\rho$. In case of large corrections, say $|a_\text{nf}+b_\text{cs}|\gtrsim0.003$, the results of all forward unpolarized proton-proton scattering measurements should be revised and, consequently, a new global fit of $\rho(s)$ and $\sigma_\text{tot}(s)$ should be carried out.

\section{Conclusions}

In this paper, the corrections to the analyzing powers given in Ref.\,\cite{bib:BKLST} were studied. For the experimental results already published,  Eqs. (\ref{eq:dI5}) and (\ref{eq:dR5}) allows one to evaluate with sufficient accuracy the corrections to the measured single spin-flip amplitude parameter $r_5$.

The improved expressions for $A_\text{N}(t)$ and $A_ \text{NN}(t)$ could be written in the same form as in Ref.\,\cite{bib:BKLST} (if neglecting the absorptive correction terms  $\Delta^a_\text{N}=\Delta^a_\text{NN}=0$)
\begin{widetext}
\begin{eqnarray}
  \frac{m_p}{\sqrt{-t}}\,A_\text{N}(t) & = & \frac 
  {\left[\varkappa'(1-\rho'\delta_C)-2(I_5-\delta_C R_5)\right]\,t'_c/t -2(R_5 -\rho'I_5) + \Delta^a_\text{N}\varkappa}
  {\left(t_c/t\right)^2-2(\tilde{\rho}+\delta_C)\,t_c/t+1+\tilde{\rho}^2},
  \label{eq:AN_c}  \\[0.5\baselineskip]
  A_\text{NN}(t) & = & \frac
  {-2(R_2 + \delta_CI_2)\,t'_c/t + 2(I_2 + \rho' R_2) - ( \rho'\varkappa-4R_5)\,\varkappa t_c/2m_p^2  + \Delta^a_\text{NN}\varkappa t/m_p^2}
  {\left(t_c/t\right)^2-2(\tilde{\rho}+\delta_C)t_c/t+1+\tilde{\rho}^2},
  \label{eq:ANN_c}  
\end{eqnarray}
\end{widetext}
but with the following modification of some parameters:
\begin{eqnarray}
  t'_c/t      &=& t_c/t  + \left(r_E^2/3-B/2-\varkappa/2m_p^2\right)t_c,\hspace*{5em} \label{eq:tc'}
  \\[0.\baselineskip]
\rho'      &=& \rho + 
\left(r_E^2/3-4/\Lambda^2-\varkappa/2m_p^2-\varkappa^2/4m_p^2\right)t_c, \label{eq:rho'}
 \\[0.\baselineskip]
 \tilde{\rho}\phantom{'}     &=& \rho - \left(4/\Lambda^2-B/2\right)t_c,
 \\[0.\baselineskip]
\varkappa' &=& \varkappa/(1\!-\!\mu_pt/4m_p^2) - 2m_p^2/(s\!-\!4m_p^2).  \label{eq:kappa'}
\end{eqnarray}

The published HJET results\,\cite{bib:HJET19} were obtained using Eqs.\,(\ref{eq:AN_c})--(\ref{eq:kappa'}) without absorptive corrections.

The double spin-flip amplitude terms in (\ref{eq:AN_c}) and the term $|\varphi_t^\text{h}|^2$ in (\ref{eq:ANN_c}) were dropped off because they are negligible for the HJET experimental accuracy and, also, are comparable with the omitted corrections of order of $(m_p^2/s)^2$ and $t/s$. The $(m_p^2/s)^2$ corrections (\ref{eq:em}) to nonflip amplitudes $\varphi_{1,3}$ were neglected in Eq.\,(\ref{eq:AN_c}). It should be noted that for experimental uncertainties similar to those at HJET these corrections become noticeable if $\sqrt{s}\!\lesssim\!5\,\text{GeV}$.

The absorptive corrections are currently undetermined [no values of $a(s)$ are published yet], but once calculated may be introduced by the following substitutions:
\begin{eqnarray}
  r_E^2/3 & \to & r_E^2/3 + a_\text{nf}/t_c,  \\
  \Delta^a_\text{N\phantom{N}} & = & a_\text{sf}-a_\text{nf}, \\
  \Delta^a_\text{NN}& = & 4R_5\,\left(a_\text{sf}-a_\text{nf}\right)
  -\rho\varkappa\,\left(a_\text{df}-a_\text{nf}\right),
\end{eqnarray}
where $a_\text{df}(s)$ is the absorptive correction (\ref{eq:abs}) to the double spin-flip electromagnetic form factor. The absorptive corrections may also affect the results of determination of $\rho$ in unpolarized elastic $pp$ scattering.
\begin{acknowledgments}
The author thanks N.H.~Buttimore, B.Z.~Kopeliovich, and M.~Krelina for stimulating discussions. N.H.~Buttimore read the manuscripts and made many valuable comments. This work was supported by Brookhaven Science Associates, LLC under Contract No.~DE-AC02-98CH\,10886 with the U.S. Department of Energy.
\end{acknowledgments}


\begin{thebibliography}{99}

\bibitem{bib:ABS}
  A.~Zelenski {\em et al.},
  %``Absolute polarized H-jet polarimeter development, for RHIC,''
  \href{https://doi.org/10.1016/j.nima.2004.08.080}%
       {Nucl. Instrum. Methods Phys. Res., Sect. A {\bf 536}, 248 (2005)}.

\bibitem{bib:PSTP2017}
  A.~Poblaguev, %{\em et al.},
  E. Aschenauer, G. Atoian, K.O. Eyser, H. Huang, Y. Makdisi, W. Schmidke, A. Zelenski, I. Alekseev, and D. Svirida, 
  %``The HJET polarimeter in RHIC Run 2017,''
  %in Proceedings of 17th International Workshop on Polarized Sources, Targets, and Polarimetry (PSTP2017), Kaist, South Korea, 2017,
  \href{https://doi.org/10.22323/1.324.0022}{{\em Proc. Sci.}, {\bf PSTP2017}, 022 (2018)}.

\bibitem{bib:Convention}
  J.~Ashkin, E.~Leader, M.~L.~Marshak, J.~B.~Roberts, J.~Soffer, and G.~H.~Thomas,
  %``Convention for Spin Parameters in High-Energy Scattering Experiments,''
  \href{https://doi.org/10.1063/1.31281}{AIP Conf. Proc. {\bf 42}, 142 (1978)};
  E. Leader, in
  \href{https://doi.org/10.1017/CBO9780511524455}{{\em Spin in Particle Physics}},
  (Cambridge University Press, Cambridge, England, 2001), p.~119.

 \bibitem{bib:KL}
  B.~Z.~Kopeliovich and L.~I.~Lapidus,
  %``On the necessity of polarization experiments in colliding p p and anti-p p beams,''
  Yad, Fiz. {\bf19}, 218 (1974) [Sov. J. Nucl. Phys. {\bf19}, 114 (1974)];
  JINR-P2-72-34 [\href{https://cds.cern.ch/record/396833}{CERN-Trans-73-7}].

\bibitem{bib:BGL}
  N.~H.~Buttimore, E.~Gotsman, and E.~Leader,
  %``Spin Dependent Phenomena Induced By Electromagnetic Hadronic Interference At High-energies,''
  \href{https://doi.org/10.1103/PhysRevD.18.694}{Phys.\ Rev.\ D {\bf 18}, 694 (1978)};
  \href{https://doi.org/10.1103/PhysRevD.35.407}{{\bf 35}, 407 (1987)}.

\bibitem{bib:BKLST}
  N.~H.~Buttimore, B.~Z.~Kopeliovich, E.~Leader, J.~Soffer, and T.~L.~Trueman,
  %``The spin dependence of high-energy proton scattering,''
  \href{https://doi.org/10.1103/PhysRevD.59.114010}{Phys.\ Rev.\ D {\bf 59}, 114010 (1999)}.

\bibitem{bib:HJET06}
  H.~Okada {\em et al.},
  %``Measurement of the analyzing power in pp elastic scattering in the peak CNI region at RHIC,''
  \href{https://doi.org/10.1016/j.physletb.2006.06.008}{Phys.\ Lett.\ B {\bf 638}, 450 (2006)}.

\bibitem{bib:HJET09}
  I.~G.~Alekseev, A.~Bravar, G.~Bunce, S.~Dhawan, K.~O.~Eyser, R.~Gill, W.~Haeberli, H.~Huang, O.~Jinnouchi, {\em et al.},
  %``Measurements of single and double spin asymmetry in pp elastic scattering in the CNI region with a polarized atomic hydrogen gas jet target,''
  \href{https://doi.org/10.1103/PhysRevD.79.094014}{Phys.\ Rev.\ D {\bf 79}, 094014 (2009)}.

\bibitem{bib:STAR13}
 L.~Adamczyk {\em et al.} (STAR Collaboration),
  %``Single Spin Asymmetry $A_N$ in Polarized Proton-Proton Elastic Scattering at $\sqrt{s}=200$ GeV,''
 \href{https://doi.org/10.1016/j.physletb.2013.01.014}{Phys.\ Lett.\ B {\bf 719}, 62 (2013)}.

\bibitem{bib:AbsorptiveCorr}
  B.~Z.~Kopeliovich and M.~Krelina,
  %``Probing the Pomeron spin-flip with Coulomb-nuclear interference,''
  \href{https://arxiv.org/abs/1910.04799}{arXiv:1910.04799v3}.
 
\bibitem{bib:Cahn} 
  R.~Cahn,
  %``Coulombic - Hadronic Interference in an Eikonal Model,''
  \href{https://doi.org/10.1007/BF01475009}{ Z.\ Phys.\ C {\bf 15}, 253 (1982)}.

\bibitem{bib:Beznogikh}
  G.~G.~Beznogikh {\it et al.},
  %``The slope parameter of the differential cross-section of elastic p-p scattering in energy range 12-70 gev,''
  \href{https://doi.org/10.1016/0370-2693(69)90438-9}{Phys.\ Lett.\  {\bf 30B}, 274 (1969)}.
  

\bibitem{bib:Menon} 
  D.~A.~Fagundes, M.~J.~Menon and P.~V.~R.~G.~Silva,
  %``Bounds on the rise of total cross section from LHC7 and LHC8 data,''
  \href{https://doi.org/doi:10.1016/j.nuclphysa.2017.06.057}{Nucl.\ Phys.\ {\bf A966}, 185 (2017)}.
%  doi:10.1016/j.nuclphysa.2017.06.057
 %[arXiv:1703.07486 [hep-ph]].

\bibitem{bib:Bartenev} 
  V.~Bartenev {\em et al.},
  %``Measurement of the Slope of the Diffraction Peak for Elastic pp Scattering from 8-GeV to 400-GeV.,''
  \href{https://doi.org/10.1103/PhysRevLett.31.1088}{Phys.\ Rev.\ Lett.\  {\bf 31}, 1088 (1973)};
  %V.~Bartenev {\em et al.},
  %``Real Part of the Proton Proton Forward Scattering Amplitude from 50 GeV to 400 GeV.,''
  %Phys.\ Rev.\ Lett.\
  \href{https://doi.org/10.1103/PhysRevLett.31.1367}{{\bf 31}, 1367 (1973)}.

\bibitem{bib:Sachs}
  F.~J.~Ernst, R.~G.~Sachs and K.~C.~Wali,
  %``Electromagnetic form factors of the nucleon,''
  \href{https://doi.org/10.1103/PhysRev.119.1105}%  
  {Phys.\ Rev.\  {\bf 119}, 1105 (1960)}.

\bibitem{bib:GD}
    L.~H.~Chan, K.~W.~Chen, J.~R.~Dunning, N.~F.~Ramsey, J.~K.~Walker, and R.~Wilson,
  %``Nucleon Form Factors and Their Interpretation,''
  \href{https://doi.org/10.1103/PhysRev.141.1298}%
  {Phys.\ Rev.\  {\bf 141}, 1298 (1966)}.
  % J.~R.~Dunning {\em et al.}, 
  %, K.~W.~Chen, A.~A.~Cone, G.~Hartwig, N.~F.~Ramsey, J.~K.~Walker and R.~Wilson,
  %``Electromagnetic Structure of the Neutron and Proton,''
  %Phys.\ Rev.\ Lett.\  {\bf 13}, 631 (1964);

\bibitem{bib:PDG}
  M. Tanabashi {\em et al.} (Particle Data Group),
  \href{https://doi.org/10.1103/PhysRevD.98.030001}{Phys. Rev. D {\bf 98}, 030001 (2018)}.

\bibitem{bib:COMPETE}
  J.~R.~Cudell, V.~V.~Ezhela, P.~Gauron, K.~Kang, Yu.~V.~Kuyanov, S.~B.~Lugovsky, E.~Martynov, B.~Nicolescu, E.~A.~Razuvaev, and N. P. Tkachenko
  (COMPETE Collaboration),
  %``Benchmarks for the forward observables at RHIC, the Tevatron Run II and the LHC,''
  \href{https://doi.org/10.1103/PhysRevLett.89.201801}%
  {Phys.\ Rev.\ Lett.\  {\bf 89}, 201801 (2002)}.
 %COMPETE Collaboration, \url{http://nuclth02.phys.ulg.ac.be/compete/predictor/}.

\bibitem{bib:rM}
 G.~Lee, J.~R.~Arrington and R.~J.~Hill,
  %``Extraction of the proton radius from electron-proton scattering data,''
 \href{https://doi.org/10.1103/PhysRevD.92.013013}%
 {Phys.\ Rev.\ D {\bf 92}, 013013 (2015)}.
%  [arXiv:1505.01489 [hep-ph]].

\bibitem{bib:HJET19}
  A.~A.~Poblaguev {\it et al.},
  %``Precision Small Scattering Angle Measurements of Elastic Proton-Proton Single and Double Spin Analyzing Powers at the RHIC Hydrogen Jet Polarimeter,''
  \href{https://doi.org/10.1103/PhysRevLett.123.162001}%
  {Phys.\ Rev.\ Lett.\  {\bf 123}, 162001 (2019)}.

\end{thebibliography}
\end{document}